\documentstyle[amssymb,aps]{revtex}

\begin{document}
\draft
\title{The center-of-mass response of confined systems }
\author{L. F. Lemmens}
\address{Dept. Natuurkunde, Universiteit Antwerpen RUCA, Groenenborgerlaan 171, B2020%
\\
Antwerpen, Belgium.}
\author{F. Brosens and J. T.\ Devreese}
\address{Dept. Natuurkunde, Universiteit Antwerpen UIA, Universiteitsplein 1, B2610\\
Antwerpen, Belgium.}
\date{September 11, 1998, resubmitted December 22, 1998 to Phys. Rev. A}
\maketitle

\begin{abstract}
For confined systems of identical particles, either bosons or fermions, we
argue that the parabolic nature of the confinement potential is a
prerequisite for the non-dissipative character of the center of mass
response to a uniform probe. For an excitation in a parabolic confining
potential, the half width of the density response function depends
nevertheless quantitatively on properties of the internal degrees of
freedom, as is illustrated here for an ideal confined gas of identical
particles with harmonic interparticle interactions.
\end{abstract}


\section{Introduction}

In the present paper we study the response of a confined system of
spin-polarized identical particles if a uniform but time-dependent force is
applied. Our method is based on the Feynman--Kac functional representing the
propagator of distinguishable particles, followed by the projection of this
functional on the irreducible representations of the permutation group as
required for identical particles. The projection introduced in \cite
{FeyBen72} is applied on confined systems. Details have been given earlier 
\cite{BDLPRE97a} to derive the thermodynamical properties and the static
response functions for bosons and for fermions. An application of the method
to real systems can be found in Ref. \cite{TBLDSSC98} for rubidium.

The response function is calculated for a spatially homogeneous but
time-dependent force. It is shown that if the confinement potential is a
polynomial of order two, the motion of the center of mass does not induce
transitions between the modes representing the relative motion of the
internal degrees of freedom. This follows from the independence of these
degrees of freedom, a property known \cite{SEPRSA55} for harmonic systems.
The analogue in extended systems is the Kohn theorem \cite{KPR61} , well
known for the cyclotron resonance in an interacting electron gas. The
connection to the response of center-of-mass excitation becomes especially
relevant if the two-body interaction only depends on the distance vector
between the positions. If one of these conditions is not satisfied, we shall
show that the homogeneous external force couples to the internal degrees of
freedom. Even if both conditions are satisfied the shape of the response
function still depends on the characteristics of the internal degrees of
freedom, typically the frequency of the internal oscillation modes and the
number of particles in the well. This suggests that these characteristics
can be estimated from the shape of the response to a homogeneous
time-dependent force. For instance, monitoring the density of a confined
system whilst changing the direction of the gravitational force on the
center of mass or changing slightly the form of the confinement potential
gives information on the internal degrees of freedom and the number of
particles confined in the system.

In view of the experimental progress made \cite
{AndSc95,DavPRL95,BraPRL95,BraPRL97} on the Bose-Einstein condensed systems,
the present response property of a perfect harmonic confined system could be
relevant to check the parabolicity of the trap. Further recent experiments 
\cite{SACISWPRL98} use strong deviations from the parabolic confinement
potential to probe the time dependent behavior of the confined atoms in a
condensed phase. In this respect it seems important to know the implications
of perfect parabolicity. Of course, the response of the system to such a
local perturbation of the density would be of major importance.

The paper is organized as follows. In Sec. II the formal response theory is
developed. In Sec. III we illustrate the response properties for the exactly
soluble model of a confined gas with harmonic interparticle interactions. In
the last section a discussion and the conclusions are given.

\section{Parabolicity and the Kohn theorem}

In a $3N$ dimensional configuration space with ${\bf r}_{j}$ $\left(
j=1,2,\ldots ,N\right) $ denoting the positions of the $N$ particles, the
position of the center of mass is defined as ${\bf R}=\frac{1}{N}
\sum_{j=1}^{N}{\bf r}_{j}$. For a quadratic confinement potential the center
of mass can be introduced as follows: 
\begin{equation}
\sum_{j=1}^{N}{\bf r}_{j}^{2}=N{\bf R}^{2}+\sum_{j=1}^{N}\left( {\bf r}_{j}-%
{\bf R}\right) ^{2}.
\end{equation}

From mathematical statistics it is well known that for Gaussian
distributions the average ${\bf R}$ can take any value without affecting the
value of $\sum_{j=1}^{N}\left( {\bf r}_{j}-{\bf R}\right) ^{2}$, i.e. the
center of mass could be independent of the deviations from that center. This
remains true in a physical system when the two-body potential depends only
on the distance between particles, because $\sum_{j,l=1}^{N}V\left( {\bf r}%
_{j}-{\bf r}_{l}\right) =\sum_{j,l=1}^{N}V\left( \left( {\bf r}_{j}-{\bf R}%
\right) -\left( {\bf r}_{l}-{\bf R}\right) \right) .$ In the harmonic model
the following two-body potential is used: 
\begin{equation}
\frac{\gamma }{4}\sum_{j,l=1}^{N}\left( {\bf r}_{j}-{\bf r}_{l}\right) ^{2}=%
\frac{\gamma N}{2}\sum_{j=1}^{N}\left( {\bf r}_{j}-{\bf R}\right) ^{2}.
\end{equation}
This two-body potential allows for exact solutions of the propagator and the
projection on the symmetric or antisymmetric irreducible representation of
the permutation group can be carried out.

Introducing a homogeneous time-dependent force: 
\begin{equation}
{\bf f}\left( \tau \right) \cdot \sum_{j=1}^{N}{\bf r}_{j}=N{\bf f}\left(
\tau \right) \cdot {\bf R}
\end{equation}
the potential confining $N$ particles and the interaction of the particles
with each other and the external field are specifying the system for which
the response will be calculated. Atomic units with $\hbar =m=1$ are used
throughout this paper.

If the particles were {\sl distinguishable}, the Euclidean-time propagator $%
K_{D}\left( {\bf \bar{r}}^{\prime \prime },\tau |{\bf \bar{r}}^{\prime
},0\right) $ for this interacting system can be obtained from a
straightforward generalization of our calculation in Ref. \cite{BDLPRE97a}.
Denoting by ${\bf \bar{r}}\in {\Bbb R}^{3N}$ the set of position vectors $%
{\bf r}_{1},\ldots ,{\bf r}_{N}$, the result turns out to be: 
\begin{equation}
K_{D}\left( {\bf \bar{r}}^{\prime \prime },\tau |{\bf \bar{r}}^{\prime
},0\right) =\frac{\left. K\left( \sqrt{N}{\bf R}^{\prime \prime },\tau |%
\sqrt{N}{\bf R}^{\prime },0\right) \right| _{\Omega }^{\sqrt{N}{\bf f}\left(
\tau \right) }}{\left. K\left( \sqrt{N}{\bf R}^{\prime \prime },\tau |\sqrt{N%
}{\bf R}^{\prime },0\right) \right| _{w}}\prod_{j=1}^{N}\left. K\left( {\bf r%
}_{j}^{\prime \prime },\tau |{\bf r}_{j}^{\prime },0\right) \right| _{w},
\label{eq:Kstr}
\end{equation}
where $\left. K\left( {\bf r}_{\tau },\tau |{\bf r}_{0},0\right) \right|
_{w} $ denotes the well-known propagator of a three-dimensional harmonic
oscillator of frequency $w$: 
\begin{equation}
\left. K\left( {\bf r}_{\tau },\tau |{\bf r}_{0},0\right) \right| _{w}=\sqrt{%
\frac{w}{2\pi \sinh w\tau }}^{3}\exp \left[ -\frac{w}{2}\frac{\left( {\bf r}%
_{\tau }^{2}+{\bf r}_{0}^{2}\right) \cosh w\tau -2{\bf r}_{\tau }\cdot {\bf r%
}_{0}}{\sinh w\tau }\right] .
\end{equation}

The quantity $K\left( {\bf R}^{\prime \prime },\tau |{\bf R}^{\prime
},0\right) _{\Omega }^{\left[ {\bf f}\left( \tau \right) \right] }$ denotes
the propagator of a three-dimensional harmonic oscillator of frequency $%
\Omega $ in the presence of a driving force ${\bf f}\left( \tau \right) .$
If the driving force acts in the real time $t$, the complex time $\tau
=\beta +it$ has to be introduced, with $\beta =1/k_{B}T$ with $k_{B}$ the
Boltzmann constant and $T$ the temperature. This propagator can be obtained
in a direct way using the functional integration techniques illustrated in 
\cite{BDLPRE97a} leading to the following result: 
\begin{eqnarray}
\left. K\left( {\bf R}^{\prime \prime },\tau |{\bf R}^{\prime },0\right)
\right| _{\Omega }^{{\bf f}\left( \tau \right) } &=&\left. K\left( {\bf R}%
^{\prime \prime },\tau |{\bf R}^{\prime },0\right) \right| _{\Omega }\times
\\
&&\exp \left( -\frac{2}{\Omega }\frac{\sinh \left( \frac{1}{2}\Omega \beta
\right) }{\cosh \left( \frac{1}{2}\Omega \beta \right) }\int_{0}^{t}%
\int_{0}^{s}{\bf f}\left( s\right) \cdot {\bf f}\left( \sigma \right) \sin
\left( \Omega \left( t-s\right) \right) \sin \left( \Omega \left( s-\sigma
\right) \right) dsd\sigma \right) \times  \nonumber \\
&&\exp \left( i{\bf R}^{\prime }\cdot \int_{0}^{t}{\bf f}\left( s\right) 
\frac{\cos \left[ \Omega \left( s-t+\frac{1}{2}i\beta \right) \right] }{%
\cosh \left( \frac{1}{2}\Omega \beta \right) }ds-i{\bf R}^{\prime \prime
}\cdot \int_{0}^{t}{\bf f}\left( s\right) \frac{\cos \left[ \Omega \left(
s-t-\frac{1}{2}i\beta \right) \right] }{\cosh \left( \frac{1}{2}\Omega \beta
\right) }ds\right) .  \nonumber
\end{eqnarray}

>From the structure of the propagator $K_{D}\left( {\bf \bar{r}}^{\prime
\prime },\tau |{\bf \bar{r}}^{\prime },0\right) $ for distinguishable
particles, it is clear that projection on the representations of the
permutation group will not affect the quotient of propagators that contains
the center of mass. Therefore, the propagator $K_{I}\left( {\bf \bar{r}}%
^{\prime \prime },\tau |{\bf \bar{r}}^{\prime },0\right) $ for {\sl identical%
} particles is given by 
\begin{equation}
K_{I}\left( {\bf \bar{r}}^{\prime \prime },\tau |{\bf \bar{r}}^{\prime
},0\right) =\frac{\left. K\left( \sqrt{N}{\bf R}^{\prime \prime },\tau |%
\sqrt{N}{\bf R}^{\prime },0\right) \right| _{\Omega }^{\sqrt{N}{\bf f}\left(
\tau \right) }}{\left. K\left( \sqrt{N}{\bf R}^{\prime \prime },\tau |\sqrt{N%
}{\bf R}^{\prime },0\right) \right| _{w}}{\Bbb K}_{I}\left( {\bf \bar{r}}%
^{\prime \prime },\tau |{\bf \bar{r}}^{\prime },0\right) ,
\end{equation}
where ${\Bbb K}_{I}\left( {\bf \bar{r}}^{\prime \prime },\tau |{\bf \bar{r}}%
^{\prime },0\right) $ accounts for the permutations of the particles, and is
discussed extensively in \cite{BDLPRE97a}. Although the present derivation
has been given for the fully harmonic model (harmonic confinement and
harmonic interaction), the line of the derivation remains valid for more
general two-body interactions. Indeed, the separation between the
propagation of the center of mass and of the other degrees of freedom is
unaffected by the introduction of the two-body potential if it only depends
on the difference vectors ${\bf r}_{j}-{\bf r}_{l}.$

\section{Response to an uniform force}

Knowing the propagator of the interacting many-particle system, the density
response can be calculated in a straightforward way as the following
expectation value: 
\begin{eqnarray}
n^{{\bf f}}\left( {\bf r},t\right) &=&\frac{1}{N}\sum_{j=1}^{N}\left\langle
\delta \left( {\bf r}-{\bf r}_{j}\right) \right\rangle =\left( \frac{1}{2\pi 
}\right) ^{3}\int n_{{\bf q}}^{{\bf f}}\left( t\right) \exp \left( -i{\bf %
q\cdot r}\right) d{\bf q},  \nonumber \\
n_{{\bf q}}^{{\bf f}}\left( t\right) &=&\frac{1}{N}\sum_{j=1}^{N}\left%
\langle \exp \left( i{\bf q\cdot r}_{j}\right) \right\rangle .
\end{eqnarray}
The averages $\left\langle A\left( {\bf \bar{r}},t\right) \right\rangle $
are defined as: 
\begin{equation}
\left\langle A\left( {\bf \bar{r}},t\right) \right\rangle =\frac{1}{Z\left(
\beta \right) }\int d{\bf \bar{r}}A\left( {\bf \bar{r}}\right) K_{I}\left( 
{\bf \bar{r}},\beta +it|{\bf \bar{r},}0\right) ,
\end{equation}
where $Z\left( \beta \right) $ is the partition function for the system
without external force at inverse temperature $\beta .$ Performing this
averages means the repetitive calculation of Gaussian integrals leading to
the following result: 
\begin{eqnarray}
n_{{\bf q}}^{{\bf f}}\left( t\right) &=&n_{{\bf q}}^{{\bf f}}\left( 0\right)
\exp \left( -i{\bf q\cdot }\int_{0}^{t}{\bf f}\left( s\right) \frac{ \sin
\Omega \left( t-s\right) }{\Omega }ds\right)  \nonumber \\
n_{{\bf q}}^{{\bf f}}\left( 0\right) &=&\exp \left\{ -\frac{{\bf q}^{2}}{4N} %
\left[ \frac{N-1}{w}\frac{\cosh \frac{1}{2}w\beta }{\sinh \frac{1}{2}w\beta }
+\frac{1}{\Omega }\frac{\cosh \frac{1}{2}\Omega \beta }{\sinh \frac{1}{2}%
\Omega \beta }\right] \right\}
\end{eqnarray}
The response function $n^{{\bf f}}\left( {\bf r},t\right) $ is then obtained
by a Fourier transform leading to a Gaussian density distribution function 
\begin{equation}
n^{{\bf f}}\left( {\bf r},t\right) =\frac{1}{\sqrt{2\pi S^{2}}^{3}}\exp
\left\{ -\frac{\left[ {\bf r}+\frac{1}{\Omega }\int_{0}^{t}{\bf f}\left(
s\right) \sin \left( \Omega \left( t-s\right) \right) ds\right] ^{2}}{2S^{2}}
\right\} ,  \label{eq: densresp}
\end{equation}
in which the variance not only depends on the frequency $\Omega $ of the
confining potential, but also on the frequency $w$ of the internal degrees
of freedom: 
\begin{equation}
S^{2}=\frac{1}{2}\left( \frac{N-1}{wN}\coth \frac{w\beta }{2}+\frac{1}{%
\Omega N}\coth \frac{\Omega \beta }{2}\right) .
\end{equation}

In the absence of interparticle interactions$,$ one readily obtains $%
S^{2}|_{w=\Omega }=\left( \coth \frac{1}{2}\Omega \beta \right) /\left(
2\Omega \right) ,$ which in the low-temperature limit $\beta \rightarrow
\infty $ gives the variance of an harmonic oscillator with frequency $\Omega
.$

In the average density response 
\begin{equation}
\left\langle {\bf r}\left( t\right) \right\rangle =\int {\bf r}n^{{\bf f}%
}\left( {\bf r},t\right) d{\bf r}=-\frac{1}{\Omega }\int_{0}^{t}{\bf f}%
\left( s\right) \sin \left( \Omega \left( t-s\right) \right) ds
\end{equation}
one clearly sees the resonant structure in the presence of an oscillating
driving force ${\bf f}\left( s\right) ={\bf f}_{v}\sin vs$ with a specific
frequency $v$: 
\begin{equation}
\left\langle {\bf r}\left( t\right) \right\rangle |_{{\bf f}\left( s\right) =%
{\bf f}_{v}\sin vs}=-\frac{{\bf f}_{v}}{\Omega }\frac{\Omega \sin tv-v\sin
\Omega t}{\Omega ^{2}-v^{2}}\text{.}
\end{equation}
This expectation value of the position does not depend on the parameters of
the internal degrees of freedom, as required by the Kohn theorem.

\section{Discussion}

From the derivation of the density response to a uniform time-dependent
force, it is clear that the factorization (\ref{eq:Kstr}) is crucial to
obtain a linear relation between the density and the applied force that does
not depend on the internal degrees of freedom. Higher-order terms in the
confinement potential will spoil this relation. A term of the third degree
in the confinement potential already introduces a contribution $%
V_{conf}\thicksim 3{\bf R}\sum_{j=1}^{N}\left( {\bf r}_{j}\right) ^{2},$
that mixes the center-of-mass motion with that of the internal degrees of
freedom, and makes transitions possible in the internal degrees of freedom
leading eventually to dissipative behavior of the center-of-mass excitation.

Furthermore the Gaussian nature of the density response may allow one to
check the parabolic character of the well by measuring the mean density and
comparing it with the trap frequencies. The standard deviations of these
measurements contain information about the excitation frequencies of the
internal degrees of freedom and the number of particles contained in the
well. The response of a center-of-mass mode behaving according to the Kohn
theorem contains information about the internal degrees of freedom. Its
resonance structure reveals the parabolic quality of the well.

It should be noted that the statistics of the internal degrees of freedom
does not enter in the expressions. Therefore the response of a parabolic
well to a uniform external force does not distinguish between fermions or
bosons merely because it only applies tot the center of mass.

\section{Acknowledgments}

This work is performed in the framework of the FWO\ projects No. 1.5.729.94,
1.5.545.98, G. 0287.95, G.0071.98 and WO.073.94N (Wetenschappelijke
Onderzoeksgemeenschap over ``Laagdimensionele systemen''), the
``Interuniversitaire Attractiepolen -- Belgische Staat, Diensten van de
Eerste Minister -- Wetenschappelijke, Technische en Culturele
Aangelegenheden'', and in the framework of the BOF\ NOI 1997 projects of the
Universiteit Antwerpen. One of the authors ( F.B.) acknowledges the FWO
(Fonds voor Wetenschappelijk Onderzoek-Vlaanderen) for financial support.

\end{document}